\documentclass[aps,prl,twocolumn,nofootinbib,superscriptaddress,showpacs,floatfix,preprintnumbers]{revtex4-2}

\usepackage[dvipsnames]{xcolor}     
\definecolor{lcolor}{rgb}{0.5,0,0}
\definecolor{citcolor}{rgb}{0,0,1}
\usepackage[breaklinks,colorlinks,urlcolor=blue,citecolor=citcolor,linkcolor=lcolor,linktoc=all]{hyperref}
\usepackage{color}
\usepackage{graphicx}	
\graphicspath{{./figures/}}
\usepackage[utf8]{inputenc}
\usepackage{amsmath} \usepackage{amssymb}
\usepackage[dvipsnames]{xcolor}      
\usepackage{bm,bbm,bbold}
\usepackage{booktabs}
\usepackage{dcolumn}

\makeatletter
\def\@fnsymbol#1{\ensuremath{\ifcase#1\or *\or \dagger\or \ddagger\or
   \mathsection\or \mathparagraph\or \|\or **\or \dagger\dagger
   \or \ddagger\ddagger \else\@ctrerr\fi}}
\makeatother

\allowdisplaybreaks

\makeatletter
\g@addto@macro\bfseries{\boldmath}
\makeatother

\usepackage{tikz}
\usepackage[customcolors]{hf-tikz}
\usepackage{mciteplus}

\makeatletter 

\renewcommand\onecolumngrid{
\do@columngrid{one}{\@ne}%
\def\set@footnotewidth{\onecolumngrid}
\def\footnoterule{\kern-6pt\hrule width 1.5in\kern6pt}%
}

\renewcommand\twocolumngrid{
        \def\footnoterule{
        \dimen@\skip\footins\divide\dimen@\thr@@
        \kern-\dimen@\hrule width.5in\kern\dimen@}
        \do@columngrid{mlt}{\tw@}
}%

\makeatother    

\newcommand{\bra}[1]{\langle #1|}
\newcommand{\ket}[1]{|#1\rangle}

\newcommand{\ii}{i}

\def\wT{{\widehat T}}
\def\wj{{\widehat j}}
\def\wJ{{\widehat J}}

\def\wP{{\widehat P}}

\def\wspt{{\widehat{\cal S}}} 

\def\wjpt{{\widehat{\cal J}}} 
\def\wPhi{{\widehat{\Phi}}}

\def\wrho{{\widehat{\rho}}}
\def\wrhol{{\widehat{\rho}_{\rm LE}}}

\newcommand{\Tr}{{\rm Tr}}

\newcommand{\x}{{\rm x}}

\newcommand{\di}{{\rm d}}
\newcommand{\be}{\begin{equation}}
\newcommand{\ee}{\end{equation}}                                                                               
\def\bea{\begin{eqnarray}}
\def\eea{\end{eqnarray}}   

\begin{document}

\title{Pseudo-gauge invariant non-equilibrium density operator}

\preprint{}

\author{Francesco Becattini}
\email{becattini@fi.infn.it}
\affiliation{Universit\'a di Firenze and INFN Sezione di Firenze, Via G. Sansone 1, I-50019, 
Sesto Fiorentino (Firenze), Italy}
\author{Carlos Hoyos}
\email{hoyoscarlos@uniovi.es}
\affiliation{Departamento de F\'{\i}sica and Instituto de Ciencias y Tecnolog\'{\i}as 
Espaciales de Asturias (ICTEA), Universidad de Oviedo,c/ Leopoldo Calvo Sotelo 18, ES-33007, Oviedo, Spain}

\begin{abstract}
We obtain a form of the local thermodynamic equilibrium density operator which is invariant under pseudo-gauge
transformations of the stress-energy and the spin tensors. This operator is an excellent candidate
to describe the dynamics of a system which is assumed to achieve local equilibrium from a pseudo-gauge
invariant quantum state, a situation which is believed to occur, for instance, in nuclear collisions at
very high energy. As a consequence of pseudo-gauge invariance, the ambiguity affecting the predictions 
of mean values of observables from a local equilibrium state can be removed.
\end{abstract}

\maketitle

{\em Introduction.}\\
One of the main issues in spin physics, especially in relativistic heavy ion collisions, 
is the dependence of the theoretical calculations of spin polarization within a quantum-relativistic framework, 
on the so-called pseudo-gauge, that is the choice of a particular spin tensor contributing to the total 
angular momentum current \cite{Becattini:2018duy,Li:2020eon,Fukushima:2020ucl,Buzzegoli:2021wlg,Drogosz:2024rbd}
\footnote{It is claimed \cite{Weickgenannt:2024ibf} that results obtained in relativistic kinetic theory 
are pseudo-gauge independent. 
However, the equations of relativistic kinetic theory at some point use matching conditions requiring the
definition of an energy-momentum density and this is where an effective pseudo-gauge dependence sneaks in.}.
The dependence appears because the quantum state which is supposed to be the best approximation to describe 
a relativistic system at local thermodynamic equilibrium depends on this arbitrary choice. While different 
pseudo-gauge choices have a relatively small impact on momentum-dependent observables, they are quantitatively 
significant for the spin polarization, as shown in ref. \cite{Buzzegoli:2021wlg}.
This ambiguity in the theoretical predictions is disturbing because in most cases the actual quantum state 
of the system is definitely pseudo-gauge independent, like in relativistic heavy ion physics where the initial 
state are two colliding nuclei. In fact, pseudo-gauge dependence is introduced by our approximation of the 
quantum state with a local equilibrium density operator as it will be discussed in detail later on. It was 
believed that such a dependence is an artifact of our modeling and ideally should be removed.

In this work, we propose a solution to this problem, that is a redefinition of the local thermodynamic equilibrium
density operator which is pseudo-gauge invariant. With this state, the theoretical calculation of any observable
becomes pseudo-gauge invariant unless the observable itself transforms, e.g. the stress-energy tensor,
in which case it becomes pseudo-gauge covariant. This especially applies to the calculation of spin polarization 
in relativistic heavy ion physics but of course it is general enough to apply to any other problem where the 
ambiguity in defining the stress-energy tensor and the angular momentum current appears.


{\em Pseudo-gauge transformations and the spin tensor}.\\
Let us first review the pseudo-gauge transformations and how they affect to the stress-energy and spin tensors. 
The stress-energy tensor $\wT^{\mu\nu}$ and the spin tensor $\wspt^{\mu,\alpha\beta}$ satisfy the 
(non-)conservation equations (or Ward identities)
\footnote{Throughout this paper square brackets on indices denote anti-symmetrization, e.g. 
$\wT^{[\mu\nu]}=\frac{1}{2}(\wT^{\mu\nu}-\wT^{\nu\mu})$ and round brackets symmetrization, 
e.g. $\wT^{(\mu\nu)}=\frac{1}{2}(\wT^{\mu\nu}+\wT^{\nu\mu})$. The spin current is antisymmetric 
in the last two indices $\wspt^{\mu,\alpha\beta}=\wspt^{\mu,[\alpha\beta]}$.} :
\begin{equation}\label{eq:wardid}
    \partial_\mu \wT^{\mu\nu}=0,\quad \partial_\mu \wspt^{\mu,\alpha\beta}=-2 \wT^{[\alpha\beta]},
\end{equation}
Combining them one can construct the total angular momentum current
\begin{equation}\label{jcurr}
    \wjpt^{\mu,\alpha\beta}=x^\alpha \wT^{\mu\beta}-x^\beta \wT^{\mu\alpha}+\wspt^{\mu,\alpha\beta},
\end{equation}
whose conservation follows from \eqref{eq:wardid}.
The right hand side of \eqref{jcurr} can be thought of the combination of the orbital contribution 
and the spin contribution to the angular momentum-boost current.

The total angular momentum-boost tensor $\wJ^{\alpha\beta}$ is obtained by integrating the total 
angular momentum current over some arbitrary 3D space-like hypersurface, that is:
\be
 \wJ^{\alpha\beta} = \int_\Sigma \di \Sigma_\mu \; \wjpt^{\mu,\alpha\beta}.
\ee
Similarly, energy and momentum are obtained by integrating the stress-energy tensor. These total
charges are conserved and are invariant under improvements of the energy-momentum tensor and spin 
current \cite{Halbwachs,Hehl:1976vr,Becattini:2011ev} that  maintain the form of the equations 
\eqref{eq:wardid}. These improvements also go under the name of pseudo-gauge transformations and depend 
on an arbitrary tensor operator antisymmetric in the last two indices 
$\wPhi^{\lambda,\mu\nu}=\wPhi^{\lambda,[\mu\nu]}$, known as super-potential. The improved quantities are
\begin{subequations}\label{eq:pseudogauge}
\begin{eqnarray}
\wspt^{\mu,\alpha\beta}_{\rm imp} &=& \wspt^{\mu,\alpha\beta}-\wPhi^{\mu,\alpha\beta} , \\   
\wT^{\mu\nu}_{\rm imp} &=& \wT^{\mu\nu}+\frac{1}{2}\partial_\lambda\left( \wPhi^{\lambda,\mu\nu}
+\wPhi^{\mu,\nu\lambda}+ \wPhi^{\nu,\mu\lambda}\right) .
\end{eqnarray}
\end{subequations}
These transformations form a group, but they cannot be associated to transformations of
the field operators, whence their name pseudo-gauge. Indeed, these transformations can be extended to 
include, on the right hand side of the equation \eqref{eq:pseudogauge}, the gradient of another arbitrary 
rank 4 tensor field:
\begin{equation}\label{zilch}
    \wspt^{\mu,\alpha\beta}_{\rm imp}=\wspt^{\mu,\alpha\beta}- \wPhi^{\mu,\alpha\beta} + 
    \partial_\sigma \widehat{\Xi}^{\sigma\mu\alpha\beta},
\end{equation}
where the tensor $\widehat{\Xi}^{\sigma\mu\alpha\beta}$, referred to as Zilch tensor, is antisymmetric in pairs 
of indices $\sigma\mu$ and $\alpha\beta$. However, we will not consider in this work the largest group
of transformations and we will confine ourselves to the form in eq. \eqref{eq:pseudogauge}. As a result of 
pseudo-gauge transformations, it is commonly stated in Quantum Field Theory that the separation between 
orbital and spin angular momentum is arbitrary. The possibility of singling out a privileged 
pseudo-gauge, or an objective separation of orbital angular momentum and spin tensor has been highly debated in other fields, such as in proton spin physics \cite{Leader:2013jra}.

One can select an improvement that makes the energy-momentum symmetric. For instance, 
by choosing $\wPhi=\wspt$ one obtains an improved symmetric stress-energy tensor and, 
correspondingly, an improved vanishing spin tensor; this is the so-called Belinfante 
pseudo-gauge. More generally, suppose that we are able to find the completely transverse 
components of the spin tensor $\wspt_\perp^{\mu,\alpha\beta}$ such that:
$$
\partial_\mu \wspt_\perp^{\mu,\alpha\beta}=\partial_\alpha \wspt_\perp^{\mu,\alpha\beta}=
\partial_\beta \wspt_\perp^{\mu,\alpha\beta}=0,
$$
(the last identity is a consequence of the second). Defining $\wspt_\parallel^{\mu,\alpha\beta}
=\wspt^{\mu,\alpha\beta}-\wspt_\perp^{\mu,\alpha\beta}$, 
the pseudo-gauge transformation that symmetrizes the energy-momentum tensor is $\wPhi^{\mu,\alpha\beta}=\wspt_\parallel^{\mu,\alpha\beta}$, in which case:
\be\label{sptrans}
\wT^{[\mu\nu]}_{\rm imp}=0 \qquad \wspt^{\mu,\alpha\beta}_{\rm imp}=\wspt_\perp^{\mu,\alpha\beta}.
\ee
Although there seem to be independent orbital and spin contributions to the angular momentum, 
the spin part does not contribute to the integral of the angular momentum density, since it 
equals a total spatial derivative:
\begin{equation}\label{eq:spintensantisym}
    \wspt_\perp^{\mu,\alpha\beta}=\partial_\lambda\left(x^{ \alpha} \widehat{V}_\perp^{\lambda\mu \beta}-x^{ \beta} \widehat{V}_\perp^{\lambda\mu \alpha}\right),
\end{equation}
where
$$
\widehat{V}_\perp^{\lambda\mu \nu}=  - \frac{1}{2}\left( \wspt_\perp^{\lambda,\mu\nu}
+\wspt_\perp^{\mu,\nu\lambda}+ \wspt_\perp^{\nu,\mu\lambda}\right).
$$
%
\begin{figure}[htb]
	\includegraphics[width=0.41\textwidth, height=0.41\textwidth]{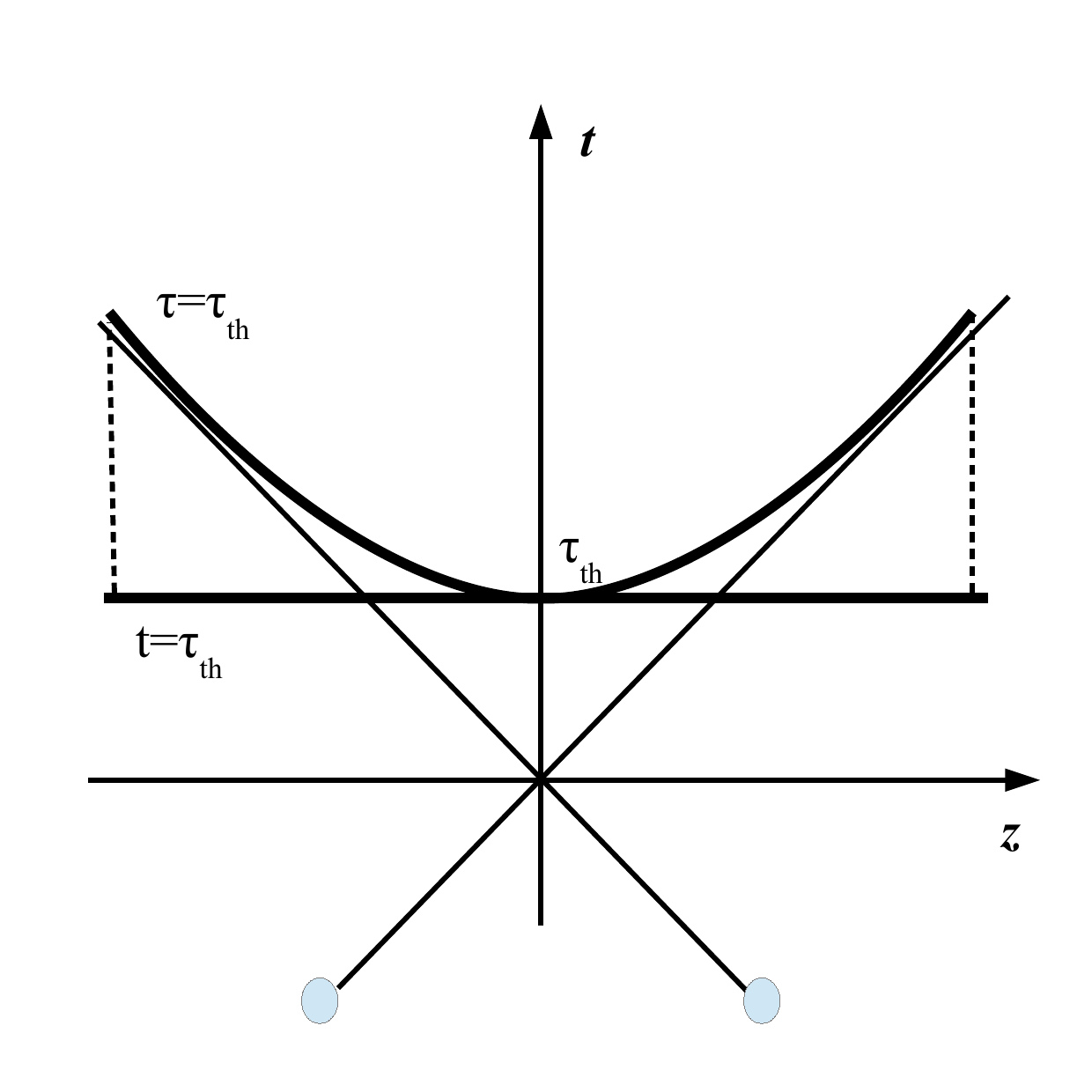}
	\caption{}
    \label{figure}
\caption{Two colliding nuclei generate a strongly interacting system which is supposed to 
achieve thermodynamic equilibrium over the hyperbola $\sqrt{t^2-z^2}=\tau=\tau_{th}$.}
\end{figure}

{\em Local equilibrium density operator and pseudo-gauge dependence}.\\
The local equilibrium density operator is obtained by maximizing the entropy $S= -\Tr (\wrho \log \wrho)$
while fixing the energy-momentum and angular momentum {\em densities} over some space-like hypersurface
$\Sigma$ (as well as, in case, conserved charge densities). The mean densities are constrained to be 
equal to the supposedly known actual values, so if $n$ is the unit vector perpendicular to the 
hypersurface $\Sigma$ we have:
\begin{align}\label{constr}
n_\mu(x) \Tr(\wrho_{LE} \wT^{\mu\nu}) &= n_\mu(x) T^{\mu\nu}, \\
n_\mu(x) \Tr(\wrho_{LE} \wjpt^{\mu,\lambda\nu}) &= n_\mu(x) {\cal J}^{\mu,\lambda\nu},
\end{align}
which, taking the equations \eqref{jcurr} and \eqref{constr} into account, implies:
$$
n_\mu(x) \Tr(\wrho_{LE} \wspt^{\mu,\lambda\nu}) =n_\mu(x) {\cal S}^{\mu,\lambda\nu}.
$$
As a result, for a generic spin tensor, the local equilibrium density operator turns out to be:
\be\label{leqd-s}
\wrho_{\rm LE} = \frac{1}{Z} \exp \left[-\int_\Sigma \di \Sigma_\mu \left(\wT^{\mu\nu} \beta_\nu 
 - \frac{1}{2} \Omega_{\lambda\nu} \wspt^{\mu,\lambda\nu} \right)\right].
\ee
where $\beta^\mu$ is the four-temperature vector and $\Omega$ is the reduced spin potential.
The operator \eqref{leqd-s} is indeed not invariant under a transformation \eqref{eq:pseudogauge} even
changing $\beta$ and $\Omega$ in any possible way. The only case where the expression \eqref{leqd-s}
becomes pseudo-gauge independent is at global thermodynamic equilibrium, where $\beta$ is a Killing 
vector, $\beta_\mu = b_\mu + \varpi_{\mu\nu} x^\nu$, with $b,\varpi$ constant and:
$$
\Omega_{\lambda\nu}=\varpi_{\lambda\nu} = -\frac{1}{2} (\partial_\lambda \beta_\nu - \partial_\nu \beta_\lambda),
$$ 
is the thermal vorticity. In this case \cite{Becattini:2012tc}, the global equilibrium density operator 
takes the following form in terms of the total momentum $\wP^\mu$ and angular momentum $\wJ^{\mu\nu}$ 
charges:
\be\label{geq}
\wrho_{\rm GE} = \frac{1}{Z} \exp \left[- b_\mu \wP^\mu + \frac{1}{2} \varpi_{\mu\nu} \wJ^{\mu\nu} 
\right].
\ee
The operator \eqref{leqd-s} is used to describe the quantum state of a relativistic fluid at some initial
time, i.e. at some hypersurface $\Sigma$. For instance, in relativistic heavy ion collisions, 
the hypersurface is the hyperbola at fixed Milne time $\tau_{th}$ (see figure \ref{figure}), the 
commonly known thermalization time of the QCD plasma. In the Heisenberg picture of quantum mechanics 
such state is fixed and is {\em the} quantum state of the system. Nevertheless, the operator \eqref{leqd-s} 
only provides an effective description; in a more rigorous approach, for a relativistic nuclear collision 
the quantum state of the system should be that of two colliding nuclei $A$ and $B$ in momentum eigenstates, 
say:
\be\label{qstate}
   \wrho_0 = \ket{P_{1A},P_{2B}} \bra{P_{1A},P_{2B}}.
\ee
One should in principle use the above state to describe particle production with the familiar formalism 
of scattering amplitudes. Unfortunately, this turns out to be practically impossible 
and the state \eqref{qstate} is effectively replaced by \eqref{leqd-s} and a subsequent hydrodynamic 
description. To make sense of how this comes about, we switch to the Schr\"odinger picture. In this 
picture the quantum state evolves with the full Hamiltonian in the Cartesian time $t$:
$$
 \wrho_S(t) \equiv \exp[-\ii \widehat H (t-t_0)] \wrho_0 \exp[\ii \widehat H (t-t_0)],
$$
while the operators become time-independent. At some later time $t=\tau_{th}$, coinciding with the
assumed local thermalization time (in Milne coordinates), we switch back to the Heisenberg picture, 
and the new Heisenberg quantum state is now $\wrho = \wrho_S(\tau_{th})$. Even though the time 
evolution operator is defined on the $t=\tau_{th}$ hypersurface shown in Figure \ref{figure}, the 
Hamiltonian operator can be equally written as an integral over the hypersurface 
$\tau=\sqrt{t^2-z^2}=\tau_{th}$ also shown in in Figure \ref{figure}:
$$
  \widehat H = \int \di^3 \x \; \wT^{00} = 
  \int_{t=\tau_{th}} \!\!\!\!\!\!\! \di \Sigma_\mu \wT^{\mu\nu} \hat t_\nu
= \int_{\tau=\tau_{th}} \!\!\!\!\!\!\! \di \Sigma_\mu \wT^{\mu\nu} \hat t_\nu,
$$
where $\hat t^\nu = (1,0,0,0)$ is the time axis unit vector. This follows form Gauss' theorem, taking 
advantage of the fact that $\partial_\mu (\wT^{\mu\nu} \hat t_\nu)=0$, provided that suitable boundary 
conditions apply at the time-like boundary of the region between the $t=\tau_{th}$ and $\tau=\tau_{th}$ 
hypersurfaces. Therefore, the new quantum state can be written as: 
\begin{align}\label{rhoevo}
\wrho = \wrho_S(\tau_{th}) = 
&\exp \left[ -\ii (\tau_{th}-t_0)\int_{\tau=\tau_{th}} \!\!\!\!\!\!\! \di \Sigma_\mu 
\wT^{\mu\nu} \hat t_\nu \right] \wrho_0 \\ \nonumber
& \times \exp \left[ +\ii (\tau_{th}-t_0)\int_{\tau=\tau_{th}} \!\!\!\!\!\!\! \di \Sigma_\mu 
\wT^{\mu\nu} \hat t_\nu \right].
\end{align}
The tacit assumption in relativistic heavy ion physics is that $\wrho$ in equation \eqref{rhoevo} 
can be very well approximated by $\wrhol$ in the equation \eqref{leqd-s} over the hypersurface $\tau=\tau_{th}$
$$
  \wrho \simeq \wrhol(\tau_{th}),
$$
for some suitable fields $\beta$ and $\Omega$. The above approximation is the essence of the hydrodynamic
model of the relativistic heavy ion collisions. However, the problem arises that while \eqref{rhoevo} is 
pseudo-gauge independent, because so are $\wrho_0$ and the Hamiltonian operator, the density 
operator \eqref{leqd-s} is not \cite{Becattini:2018duy}; this is an inconsistency of the theoretical 
model which needs to be cured.

{\em Pseudo-gauge independent density operator.}\\
Our goal is to construct a pseudo-gauge invariant operator that is a minimal generalization of 
the local equilibrium operator \eqref{leqd-s}. For this purpose, we demand that 
$\widehat\Upsilon = -\log(Z\wrhol)$ is linear in the energy-momentum and spin tensor operators 
and that it is reduced to the usual form \eqref{geq} when at global equilibrium. Therefore, we require:
\be\label{eq:rhoansatz}
\widehat\Upsilon =\int_\Sigma \di \Sigma_\mu \left(\wT^{\mu\nu} X_\nu + Y_{\lambda\nu} 
\wspt^{\mu,\lambda\nu} + Z_{\lambda\nu} \wspt^{\lambda,\mu\nu} \right),
\ee
with $X$ an unknown vector field, and $Y$ and $Z$ unknown tensor fields, with $Y$ anti-symmetric. Indeed, taking into account the anti-symmetry in the exchange of the last two 
indices of $\wspt$, \eqref{eq:rhoansatz} is the most general linear form\footnote{Note that an additional
vector field $W^{\mu}$ in the integrand, multiplied by the identity operator, can be absorbed 
in a redefinition of of $Z=\Tr (\exp[-\widehat\Upsilon])$, so the equation \eqref{eq:rhoansatz} is 
really the most general linear form.}.

Under a pseudo-gauge transformation \eqref{eq:pseudogauge} the above form gives rise to the additional
term:
\begin{align*}
\delta \widehat\Upsilon &= -\int_{\Sigma} \di \Sigma_\mu\, \left[ -\frac{1}{2} \partial_\lambda \right.
 \left(\wPhi^{\lambda,\mu\nu} + \wPhi^{\mu,\nu\lambda} + \wPhi^{\nu,\mu\lambda}\right) X_\nu  \\ \nonumber
& \left. + Y_{\lambda\nu} \wPhi^{\mu,\lambda\nu} + Z_{\lambda\nu} \wPhi^{\lambda,\mu\nu} \right] \\ \nonumber
& = -\int_{\Sigma} \di \Sigma_\mu\, \left\{ \left[ \frac{1}{2} \left(\wPhi^{\lambda,\mu\nu} + 
\wPhi^{\mu,\nu\lambda} + \wPhi^{\nu,\mu\lambda}\right) \partial_\lambda X_\nu \right.\right. \\ \nonumber 
& - \left. \frac{1}{2} \partial_\lambda \left[  \left(\wPhi^{\lambda,\mu\nu} + \wPhi^{\mu,\nu\lambda} + 
\wPhi^{\nu,\mu\lambda}\right) X_\nu \right] \right. \\ \nonumber 
& + \left. Y_{\lambda\nu} \wPhi^{\mu,\lambda\nu} + Z_{\lambda\nu} \wPhi^{\lambda,\mu\nu} \right\}.
\end{align*}
The total derivative term in the second line of the last equality 
turns out to be
the divergence of an anti-symmetric tensor in the exchange of the indices $(\mu,\lambda)$, hence it
gives rise to a 2D boundary term according to the Stokes theorem. Under suitable boundary conditions
this term vanishes and we are thus left with:
\begin{align*}
\delta \widehat\Upsilon 
& = -\int_{\Sigma} \di \Sigma_\mu\, \left[ \wPhi^{\mu,\lambda\nu} \left( Y_{\lambda\nu} - 
\frac{1}{4} \partial_\lambda X_\nu + \frac{1}{4} \partial_\nu X_\lambda \right) \right.\\
& + \left. \wPhi^{\lambda,\mu\nu} \left( Z_{\lambda\nu} + \frac{1}{2} \partial_\lambda X_\nu + 
\frac{1}{2} \partial_\nu X_\lambda \right) \right].
\end{align*}
after a rearrangement of the saturated indices. Demanding $\delta \widehat\Upsilon=0$ for arbitrary 
choices of $\wPhi$ and the spatial hypersurface leads to the following conditions:
\begin{subequations}\label{eq:solsinv}
\begin{eqnarray}
    Y_{\lambda\nu}=&\frac{1}{2} \partial_{[\lambda} X_{\nu]},\\
    Z_{\lambda\nu}=&-\partial_{(\lambda}X_{\nu)} ,
\end{eqnarray}
\end{subequations}
Now, if we want the fields $X,Y,Z$ to reproduce, under appropriate conditions, global thermodynamic 
equilibrium in the equation \eqref{eq:rhoansatz}, we have to identify $X$ with the four-temperature 
vector:
$$
   X_\nu \equiv \beta_\nu,
$$
hence, according to the equation \eqref{eq:solsinv}
\bea
 Y_{\lambda\nu} & = \frac{1}{2} \partial_{[\lambda} \beta_{\nu]} = - \frac{1}{2} \varpi_{\lambda\nu},  \\
 Z_{\lambda\nu} & = -\partial_{(\lambda}\beta_{\nu)} = - \xi_{\lambda\nu},
\eea
where $\varpi$ is the thermal vorticity and $\xi$ the thermal shear. With the above identifications, 
the pseudo-gauge invariant local equilibrium density operator becomes:
\begin{align}\label{solution}
\wrhol &= \frac{1}{Z} \exp \left[ - \int_\Sigma \di \Sigma_\mu \left( \wT^{\mu\nu} \beta_\nu - 
\frac{1}{2} \varpi_{\lambda\nu}  \wspt^{\mu,\lambda\nu} \right.\right. \\ \nonumber
& \left.\left. - \xi_{\lambda\nu} \wspt^{\lambda,\mu\nu} \right)\right].
\end{align}
At global equilibrium $\beta$ becomes a Killing vector, so the thermal shear vanishes and 
one recovers \eqref{geq}. The equation \eqref{solution} is the main result of this work.

{\em Discussion and conclusions.}\\
It is interesting to compare \eqref{solution} with the previously known form \eqref{leqd-s}.
The most important difference is the disappearance of the reduced spin potential $\Omega$ 
as an independent intensive thermodynamic variable. Indeed, demanding pseudo-gauge invariance implies 
that at local thermodynamic equilibrium the reduced spin potential must coincide with thermal 
vorticity. Of course, it is possible to {\em assume} that the equation \eqref{leqd-s}, with 
an arbitrary $\Omega$, is the quantum state of the system, but at the price of giving up
pseudo-gauge invariance. We can rephrase this by saying that the actual, pseudo-gauge invariant local 
thermodynamic equilibrium state requires $\Omega = \varpi$; a difference between the two implies 
that we are not in a local equilibrium condition.

The second important difference between \eqref{leqd-s} and \eqref{solution} is the presence of 
an additional term involving thermal shear in  \eqref{solution}; such term is not forbidden at 
local equilibrium, but it was never thought of as a possible additional term.

It can be shown that equation \eqref{solution} has the form of the density operator in the Belinfante 
pseudo-gauge after integrating by parts the under the assumption that a boundary term arising from the application 
of the Stokes theorem can be dropped, i.e. equation \eqref{solution} can be recast as:
\begin{equation}\label{belinf}
\wrhol = \frac{1}{Z} \exp \left[ - \int_\Sigma \di \Sigma_\mu \; \wT_B^{\mu\nu} \beta_\nu
\right],
\end{equation}
with:
$$
  \wT^{\mu\nu}_{B} = \wT^{\mu\nu}+\frac{1}{2}\partial_\lambda\left( \wspt^{\lambda,\mu\nu}
+\wspt^{\mu,\nu\lambda}+ \wspt^{\nu,\mu\lambda}\right) .
$$
Alternatively, one can arrive at this form by simply selecting the Belinfante pseudo-gauge in equation
\eqref{solution} with $\wspt_B^{\mu,\alpha\beta}=0$. Therefore, it turns out that the actual pseudo-gauge invariant results 
of, say, mean spin polarization \cite{Buzzegoli:2021wlg}, are those obtained thus far in the so-called 
Belinfante pseudo-gauge.

The result \eqref{solution} has also several major implications in the field of spin hydrodynamics. 
First, since the mean values of the energy-momentum tensor and spin current would transform as 
the operator themselves \eqref{eq:pseudogauge}, by using the quantum state \eqref{solution} hydrodynamics 
becomes fully covariant under pseudo-gauge transformations, like in e.g. \cite{Gallegos:2021bzp, Gallegos:2022jow}. 
However, it should be taken into account that in actual hydrodynamic calculations the field $\beta$ in 
eq. \eqref{solution} may be itself pseudo-gauge dependent because it is constrained by matching conditions 
(such as equations \eqref{constr}) involving stress-energy and spin tensor; a simple example is the definition 
of a four-velocity with the so-called Landau frame, where a stress-energy tensor is involved. Hence, whether 
spin hydrodynamics will still be relevant with the operator \eqref{solution} as an effective initial state, 
it is an open question as yet. 

Finally, we would like to point out that the above arguments cannot be extended straightforwardly
to the conserved charge currents. Let $\wj^\mu$ be a conserved current; as it is known, they also
can be modified through an improvement similar to a pseudo-gauge transformation:
\be\label{jpg}
 \wj^\mu_{\rm imp} = \wj^\mu + \partial_\lambda \widehat M^{\lambda\mu},
\ee
where $\widehat M^{\lambda\mu}=\widehat M^{[\lambda\mu]}$ is an anti-symmetric tensor. The transformation 
\eqref{jpg} preserves the conservation equation for the new current as well as the total charge under
suitable boundary conditions. The traditional local equilibrium
operator with a conserved charge is a simple extension of \eqref{leqd-s} and reads:
\be\label{leqd2}
\wrho_{\rm LE} = \frac{1}{Z} \exp \left[-\int_\Sigma \di \Sigma_\mu \left(\wT^{\mu\nu} \beta_\nu 
 - \frac{1}{2} \Omega_{\lambda\nu} \wspt^{\mu,\lambda\nu} - \zeta \wj^\mu \right)\right].
\ee
We could try to go through the same steps as for the stress-energy and spin tensor, but the
pseudo-gauge transformed term under \eqref{jpg}:
$$
\delta \widehat\Upsilon = \int_\Sigma \di \Sigma_\mu \; \zeta \partial_\lambda \widehat M^{\lambda\mu},
$$
vanishes only if $\zeta = {\rm const}$  or if $n_\mu\widehat M^{\lambda \mu}=0$. 
An example of the latter is the magnetization current that emerges from the dependence of 
the free energy on the magnetic field. 

So, in conclusion we have found a redefinition of the local thermodynamic equilibrium
density operator which is pseudo-gauge invariant under transformations mixing the stress-energy
and spin tensor. This operator may be used to effectively describe the state of a quantum
system when it is known that it must be pseudo-gauge invariant, such as in a nuclear 
collision at very high energy. The phenomenological formulae thus far obtained in literature
that should be retained as valid are those previously found in the Belinfante pseudo-gauge.
The method does not extend to conserved charge currents as well as to the widest class of pseudo-gauge
transformations involving a so-called Zilch tensor described in eq. \eqref{zilch}.

{\em Acknowledgments.}\\
This collaboration was established during the workshop ``Relativistic hydrodynamics: theory and
modern applications'' held at the Galileo Galilei Institute, Florence, Italy in 2025. 
The authors are grateful to the GGI for hospitality and for providing a stimulated environment 
during their stay. We gratefully acknowledge interesting discussions with Y. Arom, M. Buzzegoli, 
K. Fukushima, L. Gavassino, U. Heinz, I. Matthaiakakis. C.H. is 
partially supported by the Spanish Agencia Estatal de Investigaci\'on and Ministerio de Ciencia, 
Innovaci\-on y Universidades through the grants PID2021-123021NB-I00 and PID2024-161500NB-I00. F. B. is partially supported
by the project PRIN2022 Advanced Probes of the Quark Gluon Plasma funded by Ministero dell’Università 
e della Ricerca, Italy.

\end{document}